\documentstyle[prl,floats,aps]{revtex}

\begin{document}

\input{epsf.sty}

\draft

\twocolumn[\hsize\textwidth\columnwidth\hsize\csname
@twocolumnfalse\endcsname

\title{Towards an understanding of the stability properties of \\
  the 3+1 evolution equations in general relativity}

\author{
Miguel Alcubierre${}^{(1)}$,
Gabrielle Allen${}^{(1)}$,
Bernd Br\"ugmann${}^{(1)}$,
Edward Seidel${}^{(1,2)}$, and
Wai-Mo Suen${}^{(3,4)}$ \medskip
}

\address{
$^{(1)}$ Max-Planck-Institut f\"ur Gravitationsphysik,
Albert-Einstein-Institut, Am M\"{u}hlenberg 1, 14476 Golm, Germany
}
\address{
$^{(2)}$ National Center for Supercomputing Applications,
Beckman Institute, 405 N. Mathews Ave., Urbana, IL 61801
}
\address{
$^{(3)}$ Department of Physics,
Washington University, St. Louis, MO 63130
}
\address{$^{(4)}$ Physics Department,
Chinese University of Hong Kong,
Hong Kong}

\date{\today; AEI-1999-19}

\maketitle


\begin{abstract}
  We study the stability properties of the standard ADM formulation of
  the 3+1 evolution equations of general relativity through linear
  perturbations of flat spacetime.  We focus attention on modes with
  zero speed of propagation and conjecture that they are responsible
  for instabilities encountered in numerical evolutions of the ADM
  formulation. These zero speed modes are of two kinds: pure gauge
  modes and constraint violating modes.  We show how the decoupling of
  the gauge by a conformal rescaling can eliminate the problem with the
  gauge modes.  The zero speed constraint violating modes can be dealt
  with by using the momentum constraints to give them a finite speed of
  propagation.  This analysis sheds some light on the question of why
  some recent reformulations of the 3+1 evolution equations have better
  stability properties than the standard ADM formulation.
\end{abstract}


\narrowtext

\vskip2pc]


\section{Introduction}

There has been an intense effort in trying to develop Numerical
Relativity for the study of astrophysical phenomena involving black
holes and neutron stars.  Most investigations in numerical relativity
for the last 30 years have been based on the Arnowitt-Deser-Misner
(ADM)~\cite{Arnowitt62} system of evolution equations and many
important results have been obtained in spherical symmetry and
axisymmetry.  However, in the general three-dimensional (3D) case which
is needed for the simulation of realistic astrophysical systems, it has
not been possible to obtain long term stable and accurate evolutions
(although some good progress has been made, see, e.g.,
~\cite{Allen98a,Anninos96c,Bruegmann97,Cook97a}).  One might argue that
present day computational resources are still insufficient to carry out
high enough resolution 3D simulations.  However, the difficulty is
likely to be more fundamental than that.  There is no theorem
guaranteeing the well-posedness of the initial-boundary value problem
for the full ADM system.  In particular, one must consider the
possibility that free evolutions using the ADM system might be
unstable, e.g., against constraint violations in 3D.  There are also
well-known complications due to the gauge (coordinate) degrees of
freedom in the theory.  This is one of the major open problems in
numerical relativity.

Against this background of need and failure to obtain long term
stable, accurate 3D simulations in numerical relativity, in the last
decade there has been a lot of effort looking for alternative
formulations of the 3+1 equations, which can be roughly separated in
two directions.  (I) In the mathematical direction, several first
order hyperbolic formulations have been proposed, and conditions on
well-posedness of the initial-boundary value problem have been

studied~\cite{Friedrich81a,Friedrich81b,Choquet83,Friedrich85,Bona92,Bona94b,Choquet95,Frittelli95,Abrahams95c,Friedrich96,MVP95,Abrahams96a,Bona97a,Abrahams97b,Anderson97,Bona98b,Anderson99,Alcubierre99c}.
Unfortunately there is as yet no evidence that the hyperbolic
re-formulations lead to significant improvements in general 3D
numerical calculations (despite encouraging results in the spherical
symmetric case~\cite{Bona94b}).  (II) In the more ``empirical''
direction there have also been various attempts to improve stability
and accuracy by modifying the ADM system. To avoid instabilities due
to constraint violation, fully or partially constrained evolutions
have been tried and the addition of ``constraint enforcing terms''
into the ADM evolution equations has been proposed and
attempted~\cite{Detweiler87,Lai98} (cmp.~\cite{Brodbeck98}).  Methods
to better enforce gauge conditions have also been
suggested~\cite{Balakrishna96a}.  Most significant and relevant for
our present paper is an approach based on separating out the conformal
and traceless part of the ADM system, first developed by Shibata and
Nakamura~\cite{Shibata95}.  Unfortunately, the strength of the
Shibata-Nakamura approach was not widely appreciated, until Baumgarte
and Shapiro~\cite{Baumgarte99} compared the standard ADM formulation
with a modified version of the Shibata-Nakamura formulation on a
series of test cases, showing the remarkable stability properties of
the conformal-traceless (CT) system.  This has triggered much recent
research in the community, including what we are reporting here and in
a companion paper.  There also have been interesting results
connecting the conformal approach to the hyperbolic
approach~\cite{Alcubierre99c,Arbona99,Frittelli99}.

In this and a companion paper we compare the standard ADM equations to
the CT equations of Shibata-Nakamura and Baumgarte-Shapiro in
different implementations.  In the companion
paper~\cite{Alcubierre99d}, we show empirically the strength of this
system over the standard ADM equations in numerical evolutions, at
least in some of the implementations of the former set of equations.
We study in particular the CT formulations in numerical evolutions of
strongly gravitating systems (see also~\cite{Alcubierre99b}) and when
coupled to hydrodynamic evolution equations, extending previous
studies of weak fields~\cite{Baumgarte99} and of pre-determined
hydrodynamic sources~\cite{Baumgarte99b}.  The main conclusion is that
the CT formulation is more stable than the standard ADM formulation in
all cases, while it needs more resolution for a given accuracy than
ADM in some cases.

In this paper, we aim at developing a mathematical understanding of
the stability properties of the ADM and the CT equations.  Ideally
one would like to know if the different systems are well-posed.
However, the systems of equations as they stand are mixed first-second
order systems and as such are not hyperbolic in any immediate sense.
This makes a study of their well-posedness particularly difficult.
Because of this fact, we have chosen instead to study linear
perturbations of a flat background and do a Fourier analysis. We
believe that this analysis, though only valid in the linear regime,
reveals important information about the stability properties of the
different formulations.

We study in particular two types of zero speed modes that appear in
the standard ADM formulation, the ``gauge modes'' and the ``constraint
violating modes'', and what they turn into in different
implementations of the CT system.  The main result of this paper is a
conjecture that the zero speed modes are responsible for the
instabilities seen in the integration of the ADM system, and a
suggestion of how they could be handled to obtain stable evolutions.
We stress the point that we do not believe that these instabilities
are of numerical origin.  Instead, we believe that they correspond to
genuine solutions of the exact system of differential equations.  A
related analysis to the one we present here, but along different
lines, was recently carried out by Frittelli and
Reula~\cite{Frittelli99}.

In section~\ref{sec:linadm}, we study the linearized ADM equations.
In section~\ref{sec:scalar} we introduce a model problem to help us
understand the effect of the zero speed modes.  In
section~\ref{sec:conformal} we discuss the gauge modes, and in
section~\ref{sec:constraints} the constraint violating modes. In
section~\ref{sec:numerics}, numerical examples are considered. We
conclude with section~\ref{sec:conclusions}. Comments on finite
difference approximations to the linearized ADM equations can be found
in the appendix.

A final comment about the language used to describe the solutions to
the different systems of equations.  We have chosen to refer to all
solutions that satisfy the constraints as {\em physical}
solutions, and those that do not as {\em unphysical}.  According to
this criterion we will consider pure gauge solutions as physical
solutions, even if they contain no real physical information.


\section{The linearized ADM equations}
\label{sec:linadm}

Let us consider first the standard ADM evolution equations for the
spatial metric $g_{ij}$ and extrinsic curvature $K_{ij}$ which in
vacuum take the form:
\begin{eqnarray}
\left(\partial_t - \cal{L}_\beta \right) g_{ij} &=& -
2 \alpha K_{ij} , \label{eq:ADM_g} \\ \left(\partial_t - \cal{L}_\beta
\right) K_{ij} &=& - D_i D_j \alpha + \alpha \left( R_{ij} + K K_{ij}
\right. \\ \nonumber
&-& \left. 2 \; K_{im} K^m_j \right) ,
\label{eq:ADM_K}
\end{eqnarray}

\noindent with $\cal{L}_\beta$ the Lie derivative with respect to the
shift vector $\beta^i$, $\alpha$ the lapse function, $D_i$ the
covariant derivative with respect to the spatial metric, $R_{ij}$
the Ricci tensor of the 3-geometry, and $K = g^{ij} K_{ij}$.

Together with the evolution equations, one must also consider the
Hamiltonian constraint
\begin{equation}
R + K^2 - K_{ij} K^{ij} = 0 ,
\label{eq:ham}
\end{equation}

\noindent and the momentum constraints
\begin{equation}
D_j \left( K^{ij} - g^{ij} K \right) = 0 .
\label{eq:mom}
\end{equation}

Let us now take geodesic slicing $\alpha=1$ and zero shift
$\beta^i=0$, and consider as well a linear perturbation of flat space
\begin{equation}
g_{ij} = \delta_{ij} + h_{ij} ,
\end{equation}

\noindent with $h_{ij} << 1$.  The evolution equations now reduce to
\begin{eqnarray}
\partial_t  h_{ij} &=& - 2 \; K_{ij} ,
\label{eq:ADMlinear_g} \\
\partial_t  K_{ij} &=& R^{(1)}_{ij} ,
\label{eq:ADMlinear_K}
\end{eqnarray}

\noindent where the linearized Ricci tensor is given by
\begin{equation}
R^{(1)}_{ij} =  - 1/2 \; \left(
\nabla_{\rm flat}^2 h_{ij}
 - \partial_i \Gamma_j - \partial_j \Gamma_i \right).
\end{equation}

\noindent and where we have defined ($h \equiv {\rm tr} h_{ij}$)
\begin{equation}
\Gamma_i := \sum_k \partial_k h_{ik} - 1/2 \; \partial_i h ,
\end{equation}

\noindent Notice that $\Gamma_i$ is just the linearized version of
$g^{mn} \Gamma^i_{mn}$.

In the same way, we find that the linearized approximation to the
constraints is
\begin{eqnarray}
\sum_k \partial_k f_k &=& 0 , \qquad ({\rm hamiltonian})
\label{eq:ham_linear}\\
\partial_t f_i &=& 0 . \qquad ({\rm momentum})
\label{eq:mom_linear}
\end{eqnarray}

\noindent where now
\begin{equation}
f_i := \sum_k \partial_k h_{ik} - \partial_i h .
\end{equation}

The structure of the constraints is quite interesting.  They just
state that the vector $\vec{f}$ should be both divergenceless, and
time independent.  Notice that both these conditions would be
trivially satisfied if we were to choose $\vec{f}$=0, which somewhat
counter-intuitively amounts to three conditions instead of four.
Notice also that asking for a transverse ($\partial_k h_{ik}$=0) and
traceless ($h$=0) solution means that $\vec{f}$=0, so the constraints
are satisfied automatically.  This is precisely what is done when one
chooses the standard transverse-traceless (TT) gauge.

Having found the linearized evolution equations, we now proceed to do
a Fourier analysis.  Without loss of generality (but see
Appendix), we can take the plane waves to be moving in the $x$
direction.  The result for any other direction can be recovered by a
simple tensor rotation later.  We then assume that we have a solution
of the form

\begin{eqnarray}
h_{ij} &=& \hat{h}_{ij} \;  e^{i(\omega t - k x)} , \\
K_{ij} &=& \hat{K}_{ij} \;  e^{i(\omega t - k x)} .
\label{eq:spect}
\end{eqnarray}

Equation~(\ref{eq:ADMlinear_g}) implies
\begin{equation}
\hat{K}_{ij} = - \left( i \omega /2 \right)\; \hat{h}_{ij} .
\end{equation}

Substituting this into Eq.~(\ref{eq:ADMlinear_K}) we find
\begin{equation}
\omega^2 \hat{\bf h} = k^2 \; M \hat{\bf h} ,
\end{equation}

\noindent where we have defined the six-dimensional vector
\begin{equation}
\hat{\bf h} := \left( \hat{h}_{xx}, \hat{h}_{yy}, \hat{h}_{zz},
\hat{h}_{xy},
\hat{h}_{xz}, \hat{h}_{yz} \right) ,
\end{equation}

\noindent and the matrix

\begin{equation}
M = \left(
\begin{array}{cccccc}
0 & 1 & 1 & 0 & 0 & 0 \\
0 & 1 & 0 & 0 & 0 & 0 \\
0 & 0 & 1 & 0 & 0 & 0 \\
0 & 0 & 0 & 0 & 0 & 0 \\
0 & 0 & 0 & 0 & 0 & 0 \\
0 & 0 & 0 & 0 & 0 & 1
\end{array}
\right).
\label{eq:ADMmatrix}
\end{equation}

One can also find that the constraints~(\ref{eq:ham_linear})
and~(\ref{eq:mom_linear}) reduce to the three (not four!) conditions
\begin{eqnarray}
&\hat{h}_{yy} + \hat{h}_{zz} = 0 \; ,& \label{eq:cons1} \\
&\hat{h}_{xy} = 0 \; ,& \label{eq:cons2} \\
&\hat{h}_{xz} = 0 \; ,& \label{eq:cons3}
\end{eqnarray}
where the first one of these equations results from both the
hamiltonian constraint and the $x$ component of the momentum
constraint, and the last two result from the $y$ and $z$ components
of the momentum constraint respectively.

It is straightforward to calculate the eigenvalues $\lambda$ and
eigenvectors of the matrix $M$.  They turn out to be

\begin{itemize}

\item $\lambda = 0$, with corresponding eigenvectors
\begin{eqnarray}
v_1 &=& \left( 1,0,0,0,0,0 \right) , \\
v_2 &=& \left( 0,0,0,1,0,0 \right) , \\
v_3 &=& \left( 0,0,0,0,1,0 \right) .
\end{eqnarray}

\item $\lambda = 1$, with corresponding eigenvectors
\begin{eqnarray}
v_4 &=& \left( 2,1,1,0,0,0 \right) , \\
v_5 &=& \left( 0,1,-1,0,0,0 \right) , \\
v_6 &=& \left( 0,0,0,0,0,1 \right) .
\end{eqnarray}

\end{itemize}

What sort of solutions do the different eigenvectors represent?  There
are four different types of solutions:

\begin{enumerate}

\item Two modes that satisfy all the constraints that
travel with the speed of light ($\lambda = 1$) represented by the
transverse-traceless vectors $v_5$ and $v_6$.

\item One mode that violates both the hamiltonian and the $x$
component of the momentum constraints (compare with
Eq.~(\ref{eq:cons1})), that also travels with the speed of light
($\lambda = 1$) represented by the vector $v_4$.

\item Two transverse modes that violate only the momentum constraints
(compare with Eqs.~(\ref{eq:cons2}) and~(\ref{eq:cons3})), and
``travel'' with speed zero ($\lambda = 0$) represented by the vectors
$v_2$ and $v_3$.

\item One mode that satisfies all the constraints that also has speed
zero ($\lambda = 0$) represented by the vector $v_1$.

\end{enumerate}

The three constraint satisfying modes are clearly physical solutions.
Of these, the two transverse-traceless traveling modes ($v_5$ and
$v_6$) correspond to the standard gravitational waves.  What is the
remaining physical mode $v_1$?  The only possibility is for it to be a
pure gauge mode. To see that this is indeed true all we need to check
is that it corresponds to a solution for which the 4-curvature Riemann
tensor vanishes.  For this we start from the Gauss-Codazzi relations,
which to first order are
\begin{eqnarray}
{}^{(4)}R^m_{ijk} &=& {}^{(3)}R^m_{ijk} \; ,
\label{eq:GC1} \\
{}^{(4)}R^0_{ijk} &=& \partial_k K_{ij} - \partial_j K_{ik} \;
\label{eq:GC2} \\
{}^{(4)}R^0_{i0k} &=& - \partial_t K_{ik} \; ,
\label{eq:GC3}
\end{eqnarray}

Now, the fact that $v_1$ has dependence only on $x$ (by construction),
and has a component corresponding only to $h_{xx}$ implies that the
r.h.s. of~(\ref{eq:GC2}) vanishes and hence ${}^{(4)}R^0_{ijk}$=0.
Also, since this mode has zero speed, it corresponds to a mode for
which $\partial_t K_{ik}$=0 which in turn means that
${}^{(4)}R^0_{i0k}$=0.  Finally, it is not difficult to see that for a
solution that depends only on $x$ and for which only $h_{xx}$ is
non-zero, the 3-curvature vanishes as well, which tells us that
${}^{(4)}R^m_{ijk}$=0. The 4-Riemann is then identically zero, so the
mode $v_1$ is a pure gauge mode.

The presence of the zero speed modes ($v_1$, $v_2$ and $v_3$) is
troublesome: They do not represent non-evolving modes as one might
think at first sight, but rather they represent modes that annihilate
the Ricci tensor.  As such, they correspond to solutions for which the
extrinsic curvature remains constant in time, and the metric functions
grow linearly (the linearly growing gauge modes have been studied
before in~\cite{Alcubierre94b,Matzner96a}).  With the full non-linear
ADM equations, this linear growth is likely to lead to an instability.

In the next section we use a simple model problem to show how zero
speed modes can indeed become unstable in the presence of
non-linear terms.


\section{Zero speed modes: a model problem}
\label{sec:scalar}

To understand the effects of zero speed modes on stability, we study
the simple case of the one-dimensional wave equation with a non-linear
source term $F$:
\begin{equation}
{\partial_t}^2 \phi - \epsilon \; {\partial_x}^2 \phi = \delta F(\phi,
\partial_t \phi, \partial_t \phi) .
\label{eq:scalar}
\end{equation}

We investigate the stability of the system for different values of
$\epsilon$ and $\delta$.  We will call the system unstable if the
magnitude of $\phi$ grows faster than exponential in time at any fixed
value of $x$, and stable otherwise.

For $\delta=0$, but $\epsilon$ not equal to 0, we have the usual wave
equation.  A Fourier decomposition of the form used in the last
section reveals two eigen-modes with propagation speeds $\lambda = \pm
\sqrt {\epsilon}$.  The system is stable for all values of $\epsilon$
including zero, if there is no source term ($\delta=0$).  With a
source term, it will still be stable for non-zero $\epsilon$, but {\it
  not} so if $\epsilon$ becomes zero.  For zero $\epsilon$, the two
propagation speeds degenerate to zero, and the system is unstable for
a general source term.  This can be shown analytically by
writing~(\ref{eq:scalar}) in first order form:

\begin{equation}
\partial_t \left(
\begin{array}{c}
\phi \\
D \\
\pi \\
\end{array}
\right) -
\left(
\begin{array}{ccc}
0 & 0 & 0 \\
0 & 0 & 1 \\
0 & \epsilon & 0 \\
\end{array}
\right) \partial_x
\left(
\begin{array}{c}
\phi \\
D \\
\pi \\
\end{array}
\right) =
\left(
\begin{array}{c}
\pi \\
0 \\
\delta F \\
\end{array}
\right)  ,
\label{eq:1st}
\end{equation}
where $D:=\partial_x \phi$ and $\pi:=\partial_t \phi $.  For
$\epsilon$ not equal to 0, the characteristic matrix (the matrix
multiplying the $\partial_x $ term above) has three independent
eigenvectors: $(1,0,0), (0,1,\sqrt\epsilon)$, and $
(0,1,-\sqrt\epsilon)$.  The eigenvector matrix and its inverse have
bounded norms.  The system is therefore strongly hyperbolic, which in
turn guarantees its stability~\cite{Courant62}.

When $\epsilon=0$, two of the eigenvectors become degenerated and the
system becomes weakly hyperbolic~\cite{Courant62}.  For $\delta=0$, the
system is still stable, with at most linear growth in $\phi$.  But for
$\delta$ non-zero and with a general source term, the system is
unstable~\cite{Courant62,Gustafsson95}.

\begin{figure}
\epsfxsize=3.4in
\epsfysize=3.4in
\epsfbox{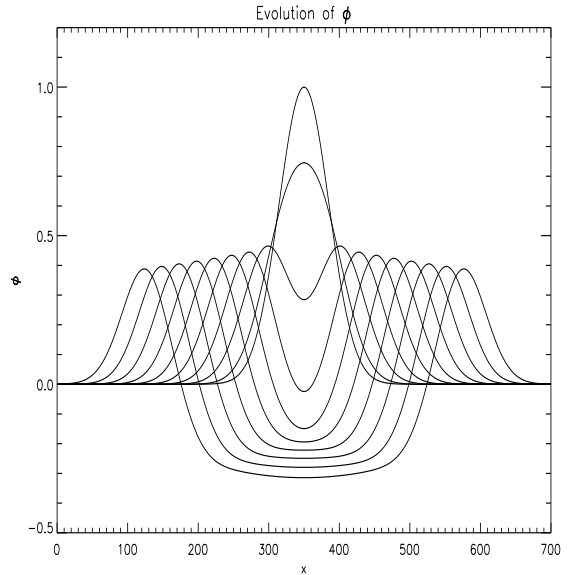}
\caption{Evolution of $\phi$ described by Eq.~(\ref{eq:scalar2}), with
$\epsilon=1$ and $\delta=-0.01$ at various times (from $t=0$ to $t=30$
in equal time intervals).}
\label{fig:ex1.1}
\end{figure}

As an example, we take  $F=\phi^2$ in (\ref {eq:scalar2}):
\begin{equation}
{\partial_t}^2 \phi - \epsilon \; {\partial_x}^2 \phi = \delta \; \phi ^2 .
\label{eq:scalar2}
\end{equation}
When $\epsilon$ is non-zero, there are no zero speed modes and the
evolution is stable.  In Fig.~\ref{fig:ex1.1} we show the evolution
of $\phi$ at various times (from $t=0$ to $t=30$ in equal time intervals)
for the case of $\epsilon=1$ and $\delta=-0.01$ (the initial data is
a Gaussian wave packet).  This evolution is very similar to that of a
non-linear gravitational plane wave (see~\cite{Anninos94d}).

\begin{figure}
\epsfxsize=3.4in
\epsfysize=3.4in
\epsfbox{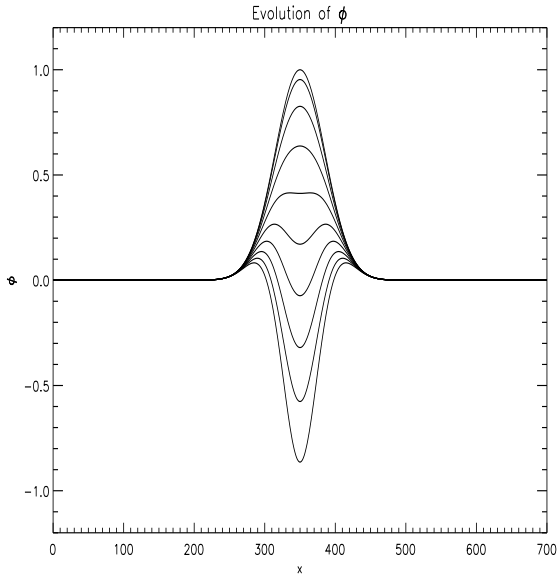}
\caption{Evolution of $\phi$ described by Eq.~(\ref {eq:scalar}), with
$\epsilon=0$ and $\delta=-0.01$ at various times (from $t=0$ to $t=5$
in equal time intervals).}
\label{fig:ex1.2}
\end{figure}

Next we tune $\epsilon$ down to zero in Eq.~(\ref{eq:scalar2}).  The
propagation speed of the eigenmodes becomes zero.  The initial
Gaussian profile now does not propagate, instead it decreases in
amplitude initially, becomes negative and eventually blows up.  See
Fig.~\ref{fig:ex1.2} for the evolution up to $t=5$, with the same
initial data as before.  In fact, this system is simple enough to be
solved exactly.  One can show that, at given value of $x$, the
solution blows up as $-1/(t-c(x))^2$, where $c$ is a constant
depending on the initial value of $\phi$ at that point.  From this it
is clear that $\phi$ in fact becomes infinite after a finite time.

We have studied examples with different source terms and have seen similar
behavior, namely the systems become unstable when $\epsilon$ goes to
zero.  To relate more directly this scalar field instability to the
ADM equations, we insert variable parameters ($\epsilon$'s) into the
linearized ADM system studied in the previous section.  We examine
the case in which the matrix $M$ (in Eq.~(\ref{eq:ADMmatrix})) contains
variable parameters $\epsilon_1 , \epsilon_2$ and $\epsilon_3$:

\begin{equation}
M_\epsilon = \left(
\begin{array}{cccccc}
\epsilon _1 & 1 & 1 & 0 & 0 & 0 \\
0 & 1 & 0 & 0 & 0 & 0 \\
0 & 0 & 1 & 0 & 0 & 0 \\
0 & 0 & 0 & \epsilon _2 & 0 & 0 \\
0 & 0 & 0 & 0 & \epsilon _3 & 0 \\
0 & 0 & 0 & 0 & 0 & 1
\end{array}
\right).
\label{eq:ADMmatrix2}
\end{equation}

For non-zero (positive) $\epsilon$'s, the corresponding set of second
order differential equations has no zero speed modes.  To investigate
its stability, it is straightforward to break this second order system
into a first order system as in the scalar field study above.  It is
then easy to show that the resulting first order system is strongly
hyperbolic and hence stable.  As we turn the $\epsilon$'s to zero
(recovering the ADM system), zero speed modes appear in the second
order system and the corresponding first order system becomes only
weakly hyperbolic.  This is precisely what happened in the scalar
field example above.  We hence conjecture that the existence of zero
speed modes and the related weak hyperbolicity is at least one of the
reasons why the ADM system becomes unstable in numerical evolutions
when non-linear source terms cannot be neglected (i.e. unstable when
gravity and/or gauge effects become strong).

As an explicit example of such a transition between linear growth and
non-linear blow up in the ADM case, we note the well known case of
focusing in geodesic slicing.  It is precisely the zero speed gauge
mode discussed in the previous section the one that represents the
focusing of geodesic slicing.  In the full non-linear case this
focusing produces a coordinate singularity causing a blow up in a
finite time.

One last comment comparing different blowing up solutions is in order.
We note that the non-linear wave equation~(\ref{eq:scalar2}) described
above has solutions that blow up in a finite time even in the case of
a non-zero wave speed.  For $\epsilon=1$ and $\delta = 1$, two such
solutions are: $\phi = -6/(t-c)^2$ (with $c$ a spatial constant) and
$\phi = -4/(t^2-x^2)$.  However, these ``blowing up solutions'' are
fundamentally different from those in the zero-wave-speed case we
focused on above.  These ``blowing up solutions'' are blowing up in a
global manner, and can come into existence in our numerical evolutions
only if we choose boundary conditions that allow them.  In numerical
evolutions (at least those considered in this paper) we typically
start the evolution at a certain initial time in a compact
computational domain with a certain chosen set of boundary conditions.
The ``blowing up solutions'', which are blowing up in a global manner
can be excluded by an appropriate set of boundary conditions.  On the
other hand, in the case when $\epsilon=0$ and $\delta > 0$, the
unstable solution involves an arbitrary function of $x$.  One can see
that any initial data with positive $\phi$ will cause a local blow up,
independently of its initial profile.  It cannot be excluded by
choosing boundary conditions.  The locality of the instability is the
crux of the problem making it dangerous in numerical evolutions.

In the next sections, we focus on the zero speed modes in the case of
the Einstein equations.  We show how one can deal separately with the
gauge mode and the constraint violating modes.


\section{Dealing with the gauge mode: decoupling $K$}
\label{sec:conformal}

In trying to deal with the zero speed modes, we will first concentrate
on the pure gauge mode: the mode $v_1$, associated with $h_{xx}$ in
the analysis of section~\ref{sec:linadm}. Since this mode satisfies
all the constraints, it represents a physical solution of the
evolution equations (even if it only corresponds to a non-trivial
evolution of the coordinate system), and hence cannot be eliminated.
The most we can hope to achieve is to decouple it from the rest of the
evolution equations, so that it will be immune to possible numerical
errors, in particular those coming from the complicated Ricci tensor
terms driving the evolution.

Remarkably, such a decoupling is not difficult to achieve.
Following~\cite{Baumgarte99,Shibata95,Arbona99} we first conformally
rescale the metric in the following way
\begin{equation}
\tilde{g}_{ij} = e^{-4 \phi} g_{ij} ,
\label{eq:rescaling}
\end{equation}
with $\phi$ chosen in such a way that the rescaled metric
$\tilde{g}_{ij}$ has unit determinant,
\begin{equation}
\phi = \frac{1}{12} \log g .
\end{equation}

We also define the conformally rescaled, trace-free part of the
extrinsic curvature $K_{ij}$ as
\begin{equation}
\tilde{A}_{ij} = e^{-4 \phi} \left( K_{ij} - \frac{1}{3} g_{ij} K
\right) .
\end{equation}

The ADM equations~(\ref{eq:ADM_g}) and~(\ref{eq:ADM_K}) can now be
rewritten as the following system of 14 evolution equations
\begin{eqnarray}
\left(\partial_t - \cal{L}_\beta \right) \phi &=& - \frac{\alpha}{6} K
, \\
\left(\partial_t - \cal{L}_\beta \right) \tilde{g}_{ij} &=& -2 \alpha
\tilde{A}_{ij} , \\
\left(\partial_t - \cal{L}_\beta \right) K &=& - D^i D_i \alpha
+ \alpha \left( R + K^2 \right) , \\
\left(\partial_t - \cal{L}_\beta \right) \tilde{A}_{ij} &=& e^{-4 \phi}
\left( - D_i D_j \alpha + \alpha R_{ij} \right)^{TF} \noindent \\
&+& \alpha \left( K \tilde{A}_{ij} - 2 \tilde{A}_{il} \tilde{A}_j^l
\right) .
\end{eqnarray}
subject to the extra constraints $\tilde{g}=1$, ${\rm tr}\tilde{A} = 0$.

The hamiltonian and momentum constraints now become as
\begin{eqnarray}
&R - \tilde{A}_{ij} \tilde{A}^{ij} + 2 K^2 / 3 = 0 \label{eq:ADMC_ham}
,& \\
&\tilde{D}_j \left( \tilde{A}^{ij} - 2 \tilde{g}^{ij} K / 3 \right)
+ 6 \tilde{A}^{ij} \partial_j \phi = 0 . & \label{eq:ADMC_mom}
\end{eqnarray}

Notice that now we have separated out the ``gauge'' variables
$\{\phi,K\}$, but we haven't yet decoupled the evolution equation for
$K$ from the Ricci tensor.  This last step can be achieved by making
use of the hamiltonian constraint above.  Doing this, we can eliminate
all reference to the Ricci tensor from the evolution equation for $K$.
One can also use the hamiltonian constraint to eliminate the Ricci
scalar from the evolution equation for $\tilde{A}_{ij}$.  In fact, one
can consider adding an arbitrary multiple of the hamiltonian
constraint to this equation. We will then consider the evolution
equations
\begin{eqnarray}
\left(\partial_t - \cal{L}_\beta \right) \phi &=& - \frac{\alpha}{6} K
, \\
\left(\partial_t - \cal{L}_\beta \right) \tilde{g}_{ij} &=& -2 \alpha
\tilde{A}_{ij} , \\
\left(\partial_t - \cal{L}_\beta \right) K &=& - D^i D_i \alpha
+ \alpha \left( \tilde{A}_{ij} \tilde{A}^{ij} + \frac{1}{3} K^2 \right)
, \\
\left(\partial_t - \cal{L}_\beta \right) \tilde{A}_{ij} &=& e^{-4 \phi}
\left(
 - D_i D_j \alpha + \alpha R_{ij} \right)^{TF}  \nonumber \\
&+& \frac{\alpha}{3} \sigma \; \tilde{g}_{ij}
\left(R - \tilde{A}_{ij} \tilde{A}^{ij} + \frac{2}{3} K^2 \right)
\nonumber \\
&+& \alpha \left(
K \tilde{A}_{ij} - 2 \tilde{A}_{il} \tilde{A}_j^l
\right) .
\end{eqnarray}

Notice that  $\sigma$=1 will correspond to the case when the Ricci
scalar is eliminated from the evolution equation for $\tilde A_{ij}$.

As before, we will now concentrate on the case of geodesic slicing
$\alpha=1$ with zero shift $\beta^i=0$, and consider a linear
perturbation of flat space.
\begin{equation}
\tilde{g}_{ij} = \delta_{ij} + \tilde{h}_{ij} ,
\end{equation}
The evolution equations then become
\begin{eqnarray}
\partial_t \phi &=& - K / 6 , \label{eq:ADMClinear_phi} \\
\partial_t \tilde{h}_{ij} &=& -2 \tilde{A}_{ij} ,
\label{eq:ADMClinear_g} \\
\partial_t K &=& 0 ,
\label{eq:ADMClinear_K} \\
\partial_t \tilde{A}_{ij} &=& R^{(1)}_{ij} - \delta_{ij} R^{(1)}
\left( 1 - \sigma \right)/3 ,
\label{eq:ADMClinear_A}
\end{eqnarray}
with $R^{(1)}_{ij}$ the linear Ricci tensor.  Notice now that to linear
order $K$ does not evolve at all: to linear order the evolution of the
gauge variables $\{\phi,K\}$ is therefore completely trivial.  In
particular, if $K$ is chosen to be zero initially, it will remain
exactly zero: no need for any exact cancellation.

Now, quite generally the Ricci tensor can be separated into
\begin{equation}
R_{ij} = \tilde{R}_{ij} + R^{\phi}_{ij} .
\label{eq:separatedRicci}
\end{equation}
The first term above $\tilde{R}_{ij}$ is the Ricci tensor associated
with the conformal metric which to linear order is
\begin{equation}
\tilde{R}^{(1)}_{ij} = - 1/2 \; \left( \nabla_{\rm flat}^2
\tilde{h}_{ij}
 - \partial_i \tilde{\Gamma_j} - \partial_j \tilde{\Gamma}_i \right) ,
\end{equation}
with the $\tilde{\Gamma}_i$ defined just as before, but now using the
conformal metric
\begin{equation}
\tilde{\Gamma}_i := \sum_k \partial_k \tilde{h}_{ik} - 1/2 \;
\partial_i
\tilde{h} .
\label{eq:tildeGamma}
\end{equation}

The second term in~(\ref{eq:separatedRicci}) is the part of the Ricci
tensor coming from the conformal factor $\phi$ which to first order is
\begin{equation}
{R^{\phi}_{ij}}^{(1)} = - 2 \left( \partial_i \partial_j \phi
+ \delta_{ij} \nabla_{\rm flat}^2 \phi \right) .
\end{equation}

Notice that one can easily prove that
\begin{equation}
\det \tilde{g}_{ij} = 1 \quad \Rightarrow \quad \tilde{h} = 0 ,
\end{equation}
so we could in principle eliminate the second term in
Eq.~(\ref{eq:tildeGamma}).  As we will see below, this is a bad idea,
so here we will just add instead a parameter $\xi$ that will be equal
to 0 if we eliminate $\tilde{h}$, and equal to 1 if we don't (but see
the next section, where the $\tilde \Gamma$'s are promoted to
independent variables).  We can then rewrite the first order Ricci
tensor as
\begin{eqnarray}
R^{(1)}_{ij} &=& - 1/2 \; \left[ \nabla_{\rm flat}^2 \tilde{h}_{ij}
+ \xi \partial_i \partial_j \tilde{h} \right]  + \sum_k \partial_k
\partial_{(i} \tilde{h}_{j)k} \nonumber \\
&-& 2 \left[ \partial_i \partial_j \phi + \delta_{ij} \nabla_{\rm
flat}^2
\phi \right] .
\end{eqnarray}

Using this we can find the linearized version of the constraints
\begin{eqnarray}
\sum_{i} \partial_i \tilde{f}_i &=& 0 , \quad ({\rm hamiltonian})
\label{eq:ADMClinear_ham} \\
\partial_t \tilde{f}_i &=& 0 . \quad ({\rm momentum})
\label{eq:ADMClinear_mom}
\end{eqnarray}
where now
\begin{equation}
\tilde{f}_i := \sum_j \partial_j \tilde{h}_{ij} - 8 \partial_i \phi .
\end{equation}

As before, having found the linearized form of the evolution
equations, we will proceed to make a Fourier analysis of the system.
We then assume that we have a solution of the form
\begin{eqnarray}
\phi &=& \hat{\phi} \; e^{i(\omega t - k x)} , \\
\tilde{h}_{ij} &=& \hat{h}_{ij} \;  e^{i(\omega t - k x)} , \\
K &=& \hat{K} \;  e^{i(\omega t - k x)} , \\
\tilde{A}_{ij} &=& \hat{A}_{ij} \;  e^{i(\omega t - k x)} .
\end{eqnarray}
The evolution equations for $\phi$ and $\tilde{h}_{ij}$ imply
\begin{equation}
\hat{K} = - 6 i \; \omega \hat{\phi} , \qquad \hat{A}_{ij} =
- \frac{i}{2} \; \omega \hat{h}_{ij} .
\end{equation}

Substituting this into the evolution equations for $K$ and
$\tilde{A}_{ij}$ we find
\begin{eqnarray}
\omega^2 \hat{\bf h} &=&  k^2 \; M \hat{\bf h} ,
\end{eqnarray}
where now ${\bf h}$ is a seven-dimensional vector
\begin{equation}
\hat{\bf h} := \left( \hat{\phi},  \hat{h}_{xx}, \hat{h}_{yy},
\hat{h}_{zz},
\hat{h}_{xy}, \hat{h}_{xz}, \hat{h}_{yz} \right) ,
\end{equation}
and
\begin{equation}
M = \left(
\begin{array}{ccccccc}
0      & 0      & 0      & 0      & 0 & 0 & 0 \\
m_{21} & m_{22} & m_{23} & m_{24} & 0 & 0 & 0 \\
m_{31} & m_{32} & m_{33} & m_{34} & 0 & 0 & 0 \\
m_{41} & m_{42} & m_{43} & m_{44} & 0 & 0 & 0 \\
0      & 0      & 0      & 0      & 0 & 0 & 0 \\
0      & 0      & 0      & 0      & 0 & 0 & 0 \\
0      & 0      & 0      & 0      & 0 & 0 & 1 \\
\end{array}
\right).
\label{eq:ADMCmatrix}
\end{equation}
with
\begin{eqnarray}
m_{21} &=& 8 - 16 (1-\sigma)/3 , \\
m_{22} &=& \left( \xi - 1 \right) \left( 1 - (1-\sigma)/3 \right) , \\
m_{23} &=& m_{24} = \xi - \left( \xi + 1 \right) (1-\sigma)/3 , \\
m_{31} &=& m_{41} = 4 - 16 (1-\sigma)/3 , \\
m_{32} &=& m_{42} = - \left( \xi - 1 \right)(1-\sigma)/3 , \\
m_{33} &=& m_{44} = 1 - \left( \xi + 1 \right) (1-\sigma)/3 , \\
m_{34} &=& m_{43} = - \left( \xi + 1 \right) (1-\sigma)/3 .
\end{eqnarray}

The hamiltonian and momentum constraints now reduce to the three
equations (again, not four!)
\begin{eqnarray}
\hat{h}_{xx} - 8 \hat{\phi} &=& 0 , \\
\hat{h}_{xy} &=& 0 , \\
\hat{h}_{xz} &=& 0 ,
\end{eqnarray}
where as before, the first condition results from both the hamiltonian
constraint and the $x$ component of the momentum constraints.

The eigenvalues $\lambda$ and eigenvectors of the
matrix~(\ref{eq:ADMCmatrix}) are now somewhat more complicated.  Let us
consider first the eigenvalues on their own.  They are:

\begin{itemize}

\item $\lambda = 0$, with multiplicity 3.

\item $\lambda = 1$, with multiplicity 2.

\item $\lambda = \left( \sigma - 1 + 3 \xi \sigma \pm \eta \right)/6$,
with

\begin{equation}
\eta = \left[ 1 + \sigma \left( 34 - 42 \xi \right) + \sigma^2 \left(
1 + 3 \xi \right)^2 \right]^{1/2}
\end{equation}

\end{itemize}

There are a couple of things to notice from the last two eigenvalues.
First, notice that if we take $\sigma$=0, one of these eigenvalues is
always negative, which implies the existence of an exponentially growing
mode, i.e. we have an unstable system of equations.  So we {\em must}
add some multiple of the hamiltonian constraint to the evolution
equation of $\tilde A_{ij}$.  How much we need to add will depend on
the value of $\xi$.  Moreover, with a little algebra one can also see
that taking $\xi$=0 results as well in a negative eigenvalue.  This
means that if we had decided to use the constraint $\tilde{h}$=0
($\xi$=0) in the expression for the Ricci tensor, we would again have
an unstable system of evolution equations.  A safe value for $\xi$
turns out to be $\xi$=1.  If we choose this, the characteristic
structure of the matrix~(\ref{eq:ADMCmatrix}) becomes

\begin{itemize}

\item $\lambda = 0$, with corresponding eigenvectors
\begin{eqnarray}
v_1 &=& \left( 1,8,-4,-4,0,0,0 \right) , \\
v_2 &=& \left( 0,0,0,0,1,0,0 \right) , \\
v_3 &=& \left( 0,0,0,0,0,1,0 \right) , \\
v_4 &=& \left( 0,1,0,0,0,0,0 \right) .
\end{eqnarray}

\item $\lambda = 1$, with eigenvectors
\begin{eqnarray}
v_5 &=& \left( 0,0,1,-1,0,0,0 \right) , \\
v_6 &=& \left( 0,0,0,0,0,0,1 \right) .
\end{eqnarray}

\item $\lambda = (4 \sigma - 1)/3$, with eigenvector
\begin{equation}
v_7 = \left( 0,(4 \sigma + 2),(4 \sigma - 1),(4 \sigma - 1),0,0,0
\right) .
\end{equation}

\end{itemize}

Notice the last eigenvalue $\lambda$ = \mbox{$(4 \sigma - 1)/3$} will
only be positive for $\sigma \geq 1/4$, which tells us that we must
add at least this much of the hamiltonian constraint to the evolution
equation for $\tilde A_{ij}$.  A natural choice is to take $\sigma$=1.
This corresponds to completely eliminating the Ricci scalar from this
equation, and results in the eigenvalue reducing to the physical speed
of light.

The type of solutions that the different eigenvectors represent
are:

\begin{enumerate}

\item Two physical solutions that travel with the speed of light
($\lambda = 1$) represented by the transverse-traceless vectors $v_5$
and $v_6$.

\item One mode that violates the hamiltonian constraint, the $x$
component of the momentum constraint, and the constraint $\tilde{h}=0$,
that travels with the speed equal to the square root of $(4 \sigma -
1)/3$, represented by the vector $v_7$.

\item Two modes that violate only the momentum constraints, and
``travel'' with speed zero ($\lambda = 0$) represented by the vectors
$v_2$ and $v_3$.

\item One mode that violates the hamiltonian constraint, the $x$
component of the momentum constraint, and the constraint $\tilde{h}=0$
that has speed zero ($\lambda = 0$) represented by the vector $v_4$.

\item One pure gauge mode (satisfying all the constraints)
that travels with speed zero ($\lambda = 0$) represented by the vector
$v_1$.

\end{enumerate}

The structure of these eigenvalues and eigenvectors tells us in the
first place that one has to be careful in the way in which different
constraints are added to the evolution equations.  The simple
statement that one is in principle free to add multiples of
constraints to evolution equations is true only if one does not worry
about the stability of the final system.  In this case we have seen
how using blindly the constraint $\tilde{h}$=0 to simplify one of the
equations results in the appearance of an unstable mode, and how
neglecting to use the hamiltonian constraint in another equation also
gives rise to an unstable mode.  A similar point has also been made
in~\cite{Frittelli97a} in the context of adding multiples of the
hamiltonian constraint to the standard ADM evolution equations.

From the characteristic structure described above, we can see that we
now have four zero speed modes instead of three (assuming we do take
$\xi$=1), so the situation would seem worse than before.  Three of
these modes are constraint violating, and we will worry about them in
the next section.  What about the gauge mode?  The gauge mode is of
course still there, and it still has zero speed (as it should), but now
it is in a much more convenient form.  From looking at $v_1$ we see
that its evolution depends on the evolution equation for $\phi$, which
we have seen is trivial in the linear and non-linear case, and the
evolution of the traceless part of $\tilde{h}_{ij}$, which is also
trivial as long as the constraint ${\rm tr} \tilde{A}$=0 is satisfied
(see Eq.~(\ref{eq:ADMClinear_g})).  The important point is the
following: the fact that this mode evolves trivially is now the
consequence of the simple {\em algebraic}\/ constraint ${\rm tr}
\tilde{A}$=0, and is independent of exact cancellations in {\em
derivatives}\/ of the metric that appear in the Ricci tensor.  This
provides an easy way to control the mode: Numerically setting ${\rm tr}
\tilde{A}$ to zero after each step of the evolution ensures that the
gauge mode cannot grow.

A comment is in order here.  It has been recognized for some
time~\cite{Bona92,Bona94b,Bona97a} that gauge modes can propagate with
arbitrary speeds.  The analysis presented above shows that constraint
violating modes can do the same.  Often one does not think about these
modes because they are unphysical, and one can avoid to excite them
with an appropriate choice of initial data.  However, from a numerical
point of view, they will never really vanish and as we have just seen
they can have important consequences on the stability of our
evolutions.  Even when these modes have a real speed of propagation
(as opposed to an imaginary speed indicating an instability on the
analytic level), if that speed is larger than the speed of light they
can cause numerical instabilities if one forgets about their existence
and chooses a time step based only on the extension of the physical
light-cones.


\section{Dealing with the constraint violating modes: using the
momentum constraints}
\label{sec:constraints}

In the previous section we have seen how separating out the gauge
variables $\{\phi,K\}$ provides a way to control the zero speed gauge
mode.  This still leaves us with the zero speed constraint violating
modes to worry about.  Here we will show how those modes can be dealt
with by using the momentum constraints to modify the evolution
equations of extra first order degrees of freedom.

The idea of using the momentum constraints to modify the evolution
equations is at the core of many recent hyperbolic reformulations of
the Einstein equations~\cite{Bona92,Bona94b,Choquet95,Frittelli95}.  In
particular, the use of the momentum constraints to obtain evolution
equations for extra first order variables can be traced back to the
Bona-Mass\'{o} formulation~\cite{Bona92,Bona94b}.  Here we will follow
for simplicity the approach of Baumgarte and
Shapiro~\cite{Baumgarte99} (a very similar approach has been used
before by Shibata and Nakamura~\cite{Shibata95}).

We will again concentrate on the case of geodesic slicing
$\alpha=1$ with zero shift $\beta^i=0$, and consider a linear
perturbation of flat space.  The linearized evolution equations
were given by~(\ref{eq:ADMClinear_phi})-(\ref{eq:ADMClinear_A})).
The Ricci tensor that appears in the evolution equation for
$\tilde{A}_{ij}$
was separated as
\begin{equation}
R_{ij}^{(1)} = \tilde{R}_{ij}^{(1)} + {R^{\phi \; (1)}_{ij}} ,
\end{equation}
with
\begin{equation}
\tilde{R}^{(1)}_{ij} = - 1/2 \; \left( \nabla_{\rm flat}^2
\tilde{h}_{ij}
 - \partial_i \tilde{\Gamma}_j - \partial_j \tilde{\Gamma}_i \right) ,
\end{equation}
and
\begin{equation}
{R^{\phi \; (1)}_{ij}} = - 2 \left( \partial_i \partial_j \phi
+ \delta_{ij} \nabla_{\rm flat}^2 \phi \right) .
\end{equation}

Now, instead of writing the quantities $\tilde{\Gamma}_i$ in terms of
their definition~(\ref{eq:tildeGamma}) as we did before, we will
promote them to independent quantities, and use~(\ref{eq:tildeGamma})
only to obtain their initial values.  We will then need an evolution
equation for the $\tilde{\Gamma}_i$.  This we can obtain trivially
from~(\ref{eq:tildeGamma}) and~(\ref{eq:ADMClinear_g}):
\begin{equation}
\partial_t \tilde{\Gamma}_i = - 2 \sum_k \partial_k \tilde{A}_{ki}
+ \partial_i {\rm tr \tilde{A}} .
\end{equation}
Notice that we can use the fact that $\tilde{A}_{ij}$ is supposed to be
traceless to eliminate the last term above.  However, we still don't
know if this will turn out to be a good idea or not, so instead we
again introduce a parameter $\xi$ and write
\begin{equation}
\partial_t \tilde{\Gamma}_i = - 2 \sum_k \partial_k \tilde{A}_{ki}
 + \xi \partial_i {\rm tr \tilde{A}} .
\end{equation}

There is still one extra modification we want to make to this
evolution equation: We will add to it a multiple of the momentum
constraints~(\ref{eq:ADMClinear_mom}) to obtain
\begin{equation}
\partial_t \tilde{\Gamma}_i = 2 \left(m - 1 \right) \sum_k \partial_k
\tilde{A}_{ki} + \xi \partial_i {\rm tr \tilde{A}}  - \frac{4m}{3}
\partial_i K ,
\label{eq:ADMClinear_G}
\end{equation}
with $m$ arbitrary.  Equation~(\ref{eq:ADMClinear_G}) above is our
final evolution equation for the $\tilde{\Gamma}_i$.  Keeping the
$\tilde{\Gamma}_i$ as independent variables, we also have to remember
that we have introduced the extra constraints $\tilde{\Gamma}_i =
\sum_k \partial_k \tilde{h}_{ik}$.

For the Fourier analysis, we again consider plane waves moving along
the $x$ direction.  From the evolution equations for $\phi$ and
$\tilde{h}_{ij}$ we find
\begin{equation}
\hat{K} = - 6 i \; \omega \hat{\phi} , \qquad \hat{A}_{ij} = -
\frac{i}{2} \;
\omega \hat{h}_{ij} ,
\end{equation}
Substituting this in the evolution equations for $\tilde{\Gamma}_i$ we
obtain
\begin{eqnarray}
\hat{\Gamma}_x &=& -i k \left[ \left( m - 1 + \xi/2 \right)
\hat{h}_{xx}
\right. \nonumber \\
&+& \left. \xi/2 \left( \hat{h}_{yy} + \hat{h}_{zz} \right)
- 8 m \hat{\phi} \right] , \\
\hat{\Gamma}_y &=& - i k \left( m - 1 \right) \hat{h}_{xy} , \\
\hat{\Gamma}_z &=& - i k \left( m - 1 \right) \hat{h}_{xz} .
\end{eqnarray}
And finally, substituting all these results back into the evolution
equations for $K$ and $\tilde{A}_{ij}$ we find

\begin{eqnarray}
\omega^2 \hat{\bf h} &=&  k^2 \; M \hat{\bf h} ,
\end{eqnarray}
where ${\bf h}$ is the same as before
\begin{equation}
\hat{\bf h} := \left( \hat{\phi},  \hat{h}_{xx}, \hat{h}_{yy},
\hat{h}_{zz},
\hat{h}_{xy}, \hat{h}_{xz}, \hat{h}_{yz} \right) ,
\end{equation}
and where the matrix $M$ is now
\begin{equation}
M = \left(
\begin{array}{ccccccc}
0      & 0      & 0      & 0      & 0 & 0 & 0 \\
m_{21} & m_{22} & m_{23} & m_{24} & 0 & 0 & 0 \\
m_{31} & m_{32} & m_{33} & m_{34} & 0 & 0 & 0 \\
m_{41} & m_{42} & m_{43} & m_{44} & 0 & 0 & 0 \\
0      & 0      & 0      & 0      & m & 0 & 0 \\
0      & 0      & 0      & 0      & 0 & m & 0 \\
0      & 0      & 0      & 0      & 0 & 0 & 1 \\
\end{array}
\right).
\label{eq:ADMBSmatrix}
\end{equation}
with
\begin{eqnarray}
m_{21} &=& 8 \left( 1-2m \right) - 16 \left( 1-m \right) (1-\sigma)/3 , \\
m_{22} &=& \left( 2 m + \xi - 1 \right) \left( 1 - (1-\sigma)/3 \right) , \\
m_{23} &=& m_{24} = \xi - \left( \xi + 1 \right) (1-\sigma)/ 3 , \\
m_{31} &=& m_{41} = 4 - 16 \left( 1 - m \right) (1-\sigma)/3 , \\
m_{32} &=& m_{42} = - \left( 2 m + \xi - 1 \right) (1-\sigma)/3 , \\
m_{33} &=& m_{44} = 1 - \left( \xi + 1 \right) (1-\sigma)/3 , \\
m_{34} &=& m_{43} = - \left( \xi + 1 \right) (1-\sigma)/3 .
\end{eqnarray}

Notice that introducing the $\tilde{\Gamma}_i$ as independent variables
by itself does not change our analysis based on $M$, which is obtained
by eliminating the $\hat{\Gamma}_i$.  But the evolution equations for
the $\tilde{\Gamma}_i$ motivate the introduction of the parameter $m$,
whose effect we consider now. The eigenvalues of the
matrix~(\ref{eq:ADMBSmatrix}) turn out to be

\begin{itemize}

\item $\lambda=0$, with multiplicity 1.

\item $\lambda=m$, with multiplicity 2.

\item $\lambda=1$, with multiplicity 2.

\item $\lambda= (1/6) \left[ b  \pm \left( b^2 - c \right)^{1/2}
\right]$, with

\begin{eqnarray}
b &=& -1 + \sigma + 3 \xi \sigma + 2 m \left( 2 + \sigma \right) , \\
c &=& 36 \sigma \left( -1 + \xi + 2 m \right) .
\end{eqnarray}

\end{itemize}

The last two eigenvalues are quite complicated, so we will concentrate
for the moment on the particular case $\sigma$=1.  In that case the
eigenvalues and eigenvectors of $M$ reduce to

\begin{itemize}

\item $\lambda = 0$, with corresponding eigenvector
\begin{equation}
v_1 = \left(1,8,-4,-4,0,0,0 \right) .
\end{equation}

\item $\lambda = m$, with corresponding eigenvectors
\begin{eqnarray}
v_2 &=& \left( 0,0,0,0,1,0,0 \right) , \\
v_3 &=& \left( 0,0,0,0,0,1,0 \right) .
\end{eqnarray}

\item $\lambda = 1$, with eigenvectors
\begin{eqnarray}
v_4 &=& \left( 0,2 \xi/(2 - 2m - \xi),1,1,0,0,0\right) , \\
v_5 &=& \left( 0,0,1,-1,0,0,0 \right) , \\
v_6 &=& \left( 0,0,0,0,0,0,1 \right) .
\end{eqnarray}

\item $\lambda = 2m + \xi - 1$, with eigenvector
\begin{equation}
v_7 = \left( 0,1,0,0,0,0,0 \right) .
\end{equation}

\end{itemize}

And the type of solutions represented are:

\begin{enumerate}

\item Two physical solutions that travel with the speed of light
($\lambda = 1$) represented by the transverse-traceless vectors $v_5$
and $v_6$.

\item One mode that violates the hamiltonian constraint, the $x$
component of the momentum constraints, and the constraint $\tilde{h}=0$
that also travels with the speed of light ($\lambda = 1$) represented
by the vector $v_4$.

\item Two modes that violate only the momentum constraints, and travel
with speed $m^{1/2}$ represented by the vectors $v_2$ and
$v_3$.

\item One mode that violates the hamiltonian constraint, the $x$
component of the momentum constraints and the constraint $\tilde{h}=0$
that has speed $(2m + \xi - 1)^{1/2}$ represented by the vector $v_7$.

\item One pure gauge mode (satisfying all the constraints)
that travels with speed zero ($\lambda = 0$) represented by the vector
$v_1$.

\end{enumerate}

Notice first that all constraint violating modes have now acquired a
non-zero speed.  If we want to have all eigenvalues non-negative (and
hence all speeds real), we must have
\begin{equation}
m \geq 0 ,
\end{equation}
and
\begin{equation}
2m + \xi - 1 \geq 0  \; \Rightarrow \; m \geq \frac{1 - \xi}{2} .
\end{equation}
In particular, if we take $\xi$=0 (that is if we use the fact that
${\rm tr}\tilde{A}$=0 in the evolution equation for $\tilde{\Gamma}_i$)
then we must have $m>1/2$.  So in order to have a stable system we {\em
must}\/ add a finite multiple of the momentum constraints to the
evolution equation for $\tilde{\Gamma}_i$.  If we fail to use the
momentum constraints, the system will have an exponentially growing
mode.  This is consistent with the results of the last section, where
we didn't have the $\Gamma_i$ (which in some sense is equivalent to not
using the momentum constraints), and we found that taking $\xi$=0
resulted in an unstable system.

Notice also that if we take
\begin{equation}
m = 1 , \qquad \xi = 0 ,
\end{equation}
then we have one zero speed mode and six modes that travel with the
speed of light.  This is precisely the choice made by Baumgarte and
Shapiro in~\cite{Baumgarte99}, so the result above explains why it was
necessary in their case to add a multiple of the momentum constraints,
and also why one should expect to have only the speed of light as a
characteristic speed in their system.  In the case $m$=1, $\xi$=0, the
eigenvector $v_4$ might appear at first sight to be singular, but from
the form that the matrix $M$ takes in this particular case it is not
difficult to show that in fact $v_4$ is replaced by $(0,0,1,1,0,0,0)$
with all other eigenvectors remaining unchanged.  The only zero speed
mode left is the pure gauge mode $v_1$, but as we have seen before,
its evolution does not rely any more on exact cancellations
in the Ricci tensor.

Finally, let us consider again the case when $\sigma \neq 1$, but
now keeping $m$=1 and $\xi$=0.  In this case the eigenvalues and
eigenvectors become

\begin{itemize}

\item $\lambda = 0$, with corresponding eigenvector
\begin{equation}
v_1 = \left(1,8,-4,-4,0,0,0 \right) .
\end{equation}

\item $\lambda = 1$, with corresponding eigenvectors
\begin{eqnarray}
v_2 &=& \left( 0,0,0,0,1,0,0 \right) , \\
v_3 &=& \left( 0,0,0,0,0,1,0 \right) , \\
v_4 &=& \left( 0,-2,1,1,0,0,0\right) , \\
v_5 &=& \left( 0,0,1,-1,0,0,0 \right) , \\
v_6 &=& \left( 0,0,0,0,0,0,1 \right) .
\end{eqnarray}

\item $\lambda = \sigma$, with eigenvector
\begin{equation}
v_7 = \left( 0,1,1,1,0,0,0 \right) .
\end{equation}

\end{itemize}

We see now that depending on how large a multiple of the hamiltonian
constraint we add to the evolution equation of $\tilde A_{ij}$, we can
change the speed of propagation of the mode that represents the trace
of $\tilde h_{ij}$ (and hence the trace of $\tilde A_{ij}$).  If we do
not use the hamiltonian constraint at all ($\sigma$=0), we will again
have a zero speed unphysical mode.  However, this is not as bad as it
might seem because in practice this mode is very easy to control since
it will vanish if one imposes the algebraic constraint ${\rm tr}
\tilde{A}$=0.


\section{Numerical examples: Stability of Minkowski spacetime}
\label{sec:numerics}

To compare the stability properties of the different systems in a
simple situation we will consider the evolution of Minkowski
spacetime, with a flat initial slice, but with a non-trivial spatial
coordinate system.  Since the extrinsic curvature is zero, the
spacetime should then remain static.  Numerically, of course, the
Ricci tensor will not be exactly zero, so we can expect some
non-trivial evolution, but if the system is stable we will only have
spurious numerical noise that should propagate away.  If the system is
unstable, however, we can expect that the numerical noise will slowly
grow in amplitude.  We will be evolving the full non-linear equations,
so the initially slow growth of the numerical noise can be expected to
trigger non-linear growth at late times.

In order to obtain our initial metric, we start from the flat space
metric in spherical coordinates
\begin{equation}
dl^2 = dr^2 + r^2 d\Omega^2 \; ,
\end{equation}
with $d\Omega^2$ the solid angle element.  We then make the
following coordinate transformation
\begin{equation}
r = \tilde{r} \left(1 - a f(\tilde{r}) \right) \; ,
\end{equation}
with $0\leq a < 1$ and $f(\tilde{r})$ a smooth monotonously decreasing
function that is 1 for small $\tilde{r}$ and 0 for large $\tilde{r}$.
The particular form of the function $f$ that we will use here is a
Gaussian
\begin{equation}
f(\tilde{r}) = e^{-\tilde{r}^2/\sigma^2} \; .
\end{equation}

In terms of the new radial coordinate the metric becomes
\begin{equation}
dl^2 = g_{11} \; d \tilde{r}^2 + \tilde{r}^2 g_{22} \; d\Omega^2 \; ,
\end{equation}
with
\begin{eqnarray}
g_{11} &=& \left[ 1 - a \left( f + \tilde{r} f' \right)
\right]^2 , \\
g_{22} &=& \left( 1 - a f \right)^2 .
\end{eqnarray}

Finally, for our 3D evolutions we transform this metric to Cartesian
coordinates in the standard way,
\begin{eqnarray}
x &=& \tilde{r} \sin \theta \cos \phi , \\
y &=& \tilde{r} \sin \theta \sin \phi , \\
z &=& \tilde{r} \cos \theta .
\end{eqnarray}
So our initial metric is
\begin{eqnarray}
g_{xx} &=& \left[ x^2 g_{11} +  \left( y^2 + z^2 \right)
g_{22}\right] / \tilde r^2 , \\
g_{yy} &=& \left[ y^2 g_{11} +  \left( x^2 + z^2 \right)
g_{22}\right] / \tilde r^2 , \\
g_{zz} &=& \left[ z^2 g_{11} +  \left( x^2 + y^2 \right)
g_{22}\right] / \tilde r^2 , \\
g_{xy} &=& x y \left( g_{11} - g_{22} \right)/ \tilde r^2 , \\
g_{xz} &=& x z \left( g_{11} - g_{22} \right)/ \tilde r^2 , \\
g_{yz} &=& y z \left( g_{11} - g_{22} \right)/ \tilde r^2 .
\end{eqnarray}

We must also say something about the gauge conditions used.  For
simplicity, we will use a zero shift vector.  For the lapse we could
try geodesic slicing, but even small numerical perturbations will
cause focusing (we are evolving the full non-linear Einstein
equations).  It is better to use a slicing that can react to the
evolution and can propagate away spurious numerical waves.  Harmonic
slicing is ideal for our purposes.  It is defined via the following
evolution equation for the lapse
\begin{equation}
\partial_t \alpha = - \alpha^2 K .
\end{equation}
Since $K$ is initially set to $0$, the lapse should remain $1$ if the
evolution is exact.

Finally, a comment about boundary conditions.  We have used a very
simple `zero order extrapolation' boundary condition, that is, we
update the boundary by just copying the value of a given field from its
value one grid point in (along the normal direction to the boundary).
This condition is not very physical, nor does it allow waves to leave
the computational grid cleanly enough, but it is very robust, and can
be used with all the different formulations studied here in a stable
way (at least for the time scales under study).  Since our emphasis is
on the stability of the interior evolution, we are content with having
a stable boundary condition.  We have used more sophisticated boundary
conditions in various cases, but it is difficult to find one that will
remain stable for all the evolution systems considered.

We now present results of simulations performed with the different
systems.  The numerical method used in all these simulations was the
so-called `iterative Crank-Nicholson' (ICN) scheme with three
iterations.  We have found that three iterations are enough to obtain
a stable, second order accurate numerical scheme~\cite{Alcubierre99d}.

\begin{figure}
\epsfxsize=3.4in
\epsfysize=3.4in
\epsfbox{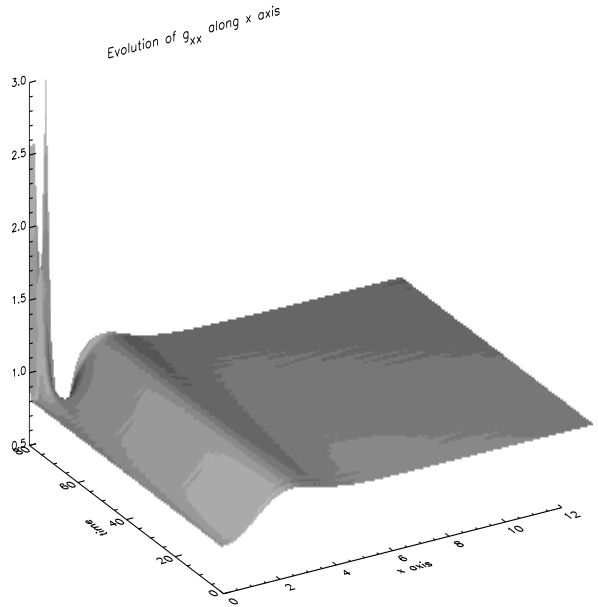}
\caption{Surface plot of $g_{xx}$ along the $x$ axis as a function of
time for the simulation using the standard ADM formulation.}
\label{fig:ADM1}
\end{figure}

\begin{figure}
\epsfxsize=3.4in
\epsfysize=3.4in
\epsfbox{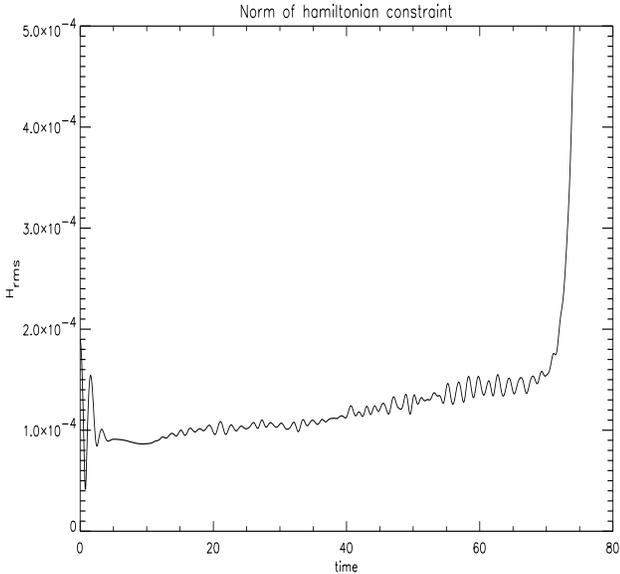}
\caption{Root mean square of the hamiltonian constraint as a function
of time for the simulation using the standard ADM formulation.}
\label{fig:ADM2}
\end{figure}

First we show the results of a simulation using the standard ADM
formulation for the case when $a$=0.1 and $\sigma$=2.  For this
simulation we used a grid with $64^3$ points and a resolution on
$\Delta x = 0.2$.  Figure~\ref{fig:ADM1} shows a surface plot of
$g_{xx}$ along the $x$ axis as a function of time.  We see that
$g_{xx}$ keeps its initial shape for some time, but at late times it
starts to fall apart near the center.  The simulation finally crashes
at $t$=79.  Figure~\ref{fig:ADM2} shows the root mean square (r.m.s.)
of the hamiltonian constraint over the numerical grid as a function of
time.  We see that for a long time there is an essentially linear
growth of the r.m.s. of the hamiltonian constraint superimposed with
small oscillations, just what we expect from the linear analysis of
the previous sections.  At late times, however, the non-linear effects
take over and we have a catastrophic blow-up, as we argued above.

\begin{figure}
\epsfxsize=3.4in
\epsfysize=3.4in
\epsfbox{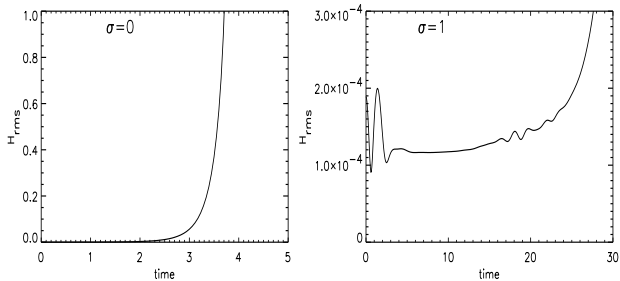}
\vspace{-1.5in}
\caption{Root mean square of the hamiltonian constraint as a function
  of time for the simulation using the standard CT formulation with
  $\xi$=0 and two different values of $\sigma$.}
\label{fig:ADMC1}
\end{figure}

Next, we show results of the conformally rescaled system of
section~\ref{sec:conformal}, using $\xi$=1, and two different values of
$\sigma$: $\sigma$=0 (no use of the hamiltonian constraint) and
$\sigma$=1 (use of the hamiltonian constraint to completely eliminate
the Ricci scalar from the evolution equation for $\tilde A_{ij}$).
From our analysis we expect the system with $\sigma$=0 to have an
exponentially growing mode and thus to be very unstable.  The
$\sigma$=1 should only have the zero speed modes and should be much
more stable (but still not completely stable).  Figure~\ref{fig:ADMC1}
shows the r.m.s. of the hamiltonian constraint for these two runs.  We
see that our predictions are indeed correct, the $\sigma$=0 run becomes
rapidly unstable and crashes at $t$=4, while the $\sigma$=1 is far more
stable and only crashes at $t$=33.

\begin{figure}
\epsfxsize=3.4in
\epsfysize=3.4in
\epsfbox{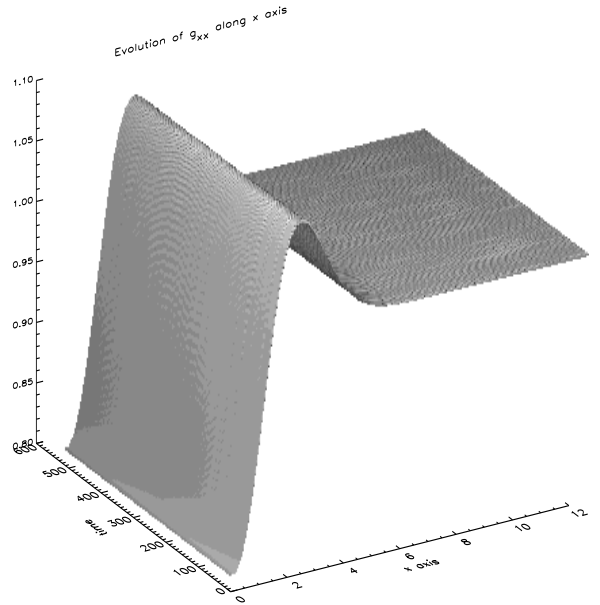}
\caption{Surface plot of $g_{xx}$ along the $x$ axis as a function of
time for the simulation using the Baumgarte-Shapiro formulation.}
\label{fig:BSSN1}
\end{figure}

\begin{figure}
\epsfxsize=3.4in
\epsfysize=3.4in
\epsfbox{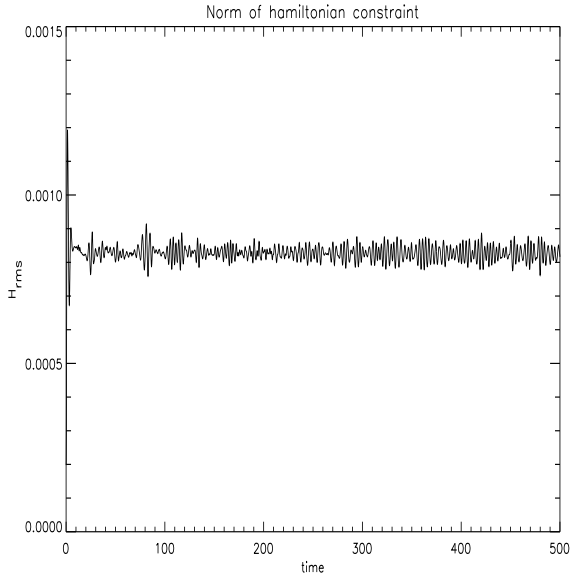}
\caption{Root mean square of the hamiltonian constraint as a function
of time for the simulation using the Baumgarte-Shapiro formulation.}
\label{fig:BSSN2}
\end{figure}

We now show the results of the choice $\sigma$=0, $m$=1, $\xi$=0 in
section~\ref{sec:constraints}, as used by
Baumgarte-Shapiro~\cite{Baumgarte99}.  We have set $tr \tilde A_{ij}$
to zero at each step as discussed above.  Figure~\ref{fig:BSSN1} shows
again a surface plot of $g_{xx}$ along the $x$ axis as a function of
time (but notice the change of scale).  The evolution now goes past
$t$=500 with no trace of an instability. Figure~\ref{fig:BSSN2} shows
the r.m.s. of the hamiltonian constraint for this run.  The
hamiltonian constraint rapidly becomes much larger than in the ADM
case at early times (by almost a factor of 10).  However, it then
stops growing and simply oscillates around a constant value, showing
again no sign of the linear growth or the blow-up that we saw for ADM.

\begin{figure}
\epsfxsize=3.4in
\epsfysize=3.4in
\epsfbox{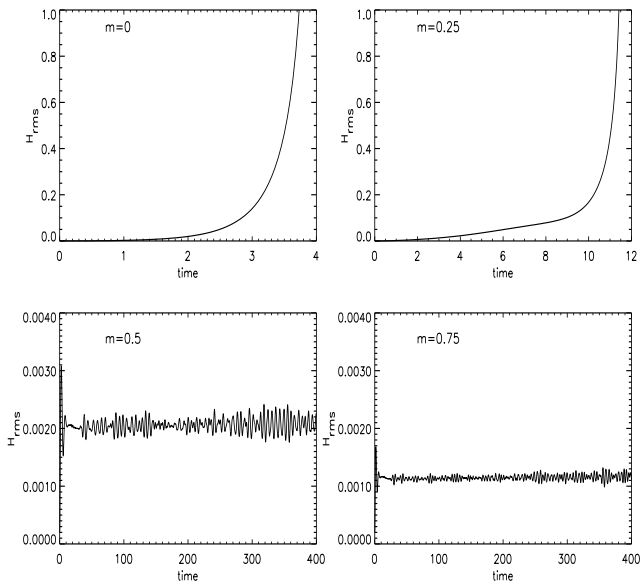}
\caption{Root mean square of the hamiltonian constraint as a function
of time for the simulation using the Baumgarte-Shapiro formulation
with different multiples of the momentum constraint added to the
evolution equation for the $\tilde \Gamma$'s (different values of the
parameter $m$).}
\label{fig:BSSN3}
\end{figure}

Finally, we show results of a series of simulations done by keeping
$\sigma$=0 and $\xi$=0, but changing the value of $m$ (the amount of
momentum constraint added to the evolution equation of the $\tilde
\Gamma$'s).  Figure~\ref{fig:BSSN3} shows the r.m.s. of the hamiltonian
constraint for runs with $m=\{0,0.25,0.5,0.75\}$ (compare with the
$m$=1 case shown above).  As expected from our analysis, we see that
the cases with $m<1/2$ rapidly become unstable.  The simulation with
$m$=0 crashes at $t$=4 while the one with $m$=0.25 crashes at
$t$=12. On the other hand, the cases with $m \geq 0.5$ remain stable
past $t$=400.

\section{Conclusions}
\label{sec:conclusions}

We have studied the stability properties of the standard ADM
formulation of general relativity based on a linear perturbation
analysis.  We focus attention on the zero speed modes.  We conjecture
that the zero speed modes can cause instabilities in evolutions of the
ADM system in its standard form.  These instabilities do not have a
numerical origin, but rather they correspond to genuine blowing-up
solutions of the differential equations.

We show that the zero speed modes come in two forms: a pure gauge mode
that satisfies all the constraints, and is therefore a legitimate
physical solution, and a series of non-physical constraint violating
modes.  We investigate the change in behavior of these modes going
from the standard ADM formulation to the conformal-traceless (CT)
systems of Shibata and Nakamura ~\cite{Shibata95} and Baumgarte and
Shapiro ~\cite{Baumgarte99}, and their derivatives.  Two features we
believe responsible for the better stability property of the conformal
systems are identified: 1. The zero speed gauge mode is governed by an
equation that is free from the complication of the Ricci tensor, thus
decoupling it from the rest of the system.  2. The constraint
violating zero speed modes, on the other hand, acquire a finite speed
of propagation due to the introduction of extra first order degrees of
freedom, and the use of the momentum constraints to modify the
evolution equations for these degrees of freedom.  We present
numerical examples to support our analysis.

We consider the study presented in this paper as a first step towards
the understanding of the stability issue in the numerical evolution of
the Einstein equations.


\acknowledgements

The authors would like to thank A. Arbona, J. Baker, C. Bona,
H. Friedrich, J. Mass\'{o}, M. Miller, A. Rendall and O. Reula
for many helpful discussions.  The research is supported
in part by NSF Phy 96-00507, NASA NCCS5-153 and NRAC grant MCA 93S025.


\appendix

\section{Finite difference approximation to the linearized ADM equations}
\label{sec:findiff}

We will consider a simple finite difference approximation to the
linearized ADM evolution equations written in second order form.  For
this we start from equations~(\ref{eq:ADMlinear_g})
and~(\ref{eq:ADMlinear_K}), and substitute one into the other to find
\begin{equation}
\partial_t^2 h_{ij} - \nabla^2_{\rm flat} h_{ij} = \partial_i
\partial_j h
 - 2 \sum_m \partial_{(i} \partial_m h_{j)m} \; .
\label{eq:linear_h2}
\end{equation}

We now construct a simple second order finite difference approximation
to this equation using standard centered differences,
\begin{eqnarray}
\partial_t^2 f & \simeq & \frac{1}{(\Delta t)^2} \delta_t^2 f^n_{\bf m}
\; , \\
\partial_i^2 f & \simeq & \frac{1}{(\Delta x)^2} \delta_i^2 f^n_{\bf m}
\; ,
\end{eqnarray}
with $f^n_{\bf m} = f(t=n \Delta t,x_i=m_i \Delta x)$ and
\begin{eqnarray}
\delta_t^2 f^n_{\bf m} &=& f^{n+1}_{\bf m} - 2 f^n_{\bf m} +
f^{n-1}_{\bf m} \; , \\
\delta_i^2 f^n_{\bf m} &=& f^n_{m_i+1} - 2 f^n_{m_i} + f^n_{m_i-1} \; ,
\end{eqnarray}

Let us now consider a plane wave solution of the form
\begin{equation}
({h_{ij}})^n_{\bf m} = \hat{h}_{ij} e^{i(n \omega \Delta t + {\bf
m}\cdot{\bf k}
\Delta x)} \; .
\end{equation}
But notice now that we allow the waves to move along any direction.
This is because even if different directions are equivalent from the
analytic point of view, they are not equivalent numerically because
the numerical grid introduces preferred directions.

If we substitute this into the finite difference approximation
to~(\ref{eq:ADMlinear_K}) we find the following equation,
\begin{equation}
\frac{2}{\rho^2} \left[ 1 - \cos(\omega \Delta t)
\right] \hat{\bf h} = \tilde{M} \hat{\bf h} \; ,
\label{eq:FDA}
\end{equation}
where $\rho := \Delta t / \Delta x$ is the Courant parameter,
$\hat{\bf h}$ is defined as before,
\begin{equation}
\hat{\bf h} := \left( \hat{h}_{xx}, \hat{h}_{yy}, \hat{h}_{zz},
\hat{h}_{xy},
\hat{h}_{xz}, \hat{h}_{yz} \right) ,
\end{equation}
and $\tilde{M}$ is the matrix

\begin{equation}
\tilde{M} = \left(
\begin{array}{cccccc}
u_y^2+u_z^2 & u_x^2       & u_x^2       & 2 s_{xy} & 2 s_{xz} & 0
 \\
u_y^2       & u_x^2+u_z^2 & u_y^2       & 2 s_{xy} & 0        & 2
s_{yz} \\
u_z^2       & u_z^2       & u_x^2+u_y^2 & 0        & 2s_{xz}  & 2
s_{yz} \\
0           & 0           & -s_{xy}     & u_z^2    & s_{yz}   & s_{xz}
\\
0           & -s_{xz}     & 0           & s_{yz}   & u_y^2    & s_{xy}
\\
-s_{yz}     & 0           & 0           & s_{xz}   & s_{xy}   & u_x^2
\end{array}
\right),
\label{eq:FDAmatrix}
\end{equation}

\noindent where we have defined
\begin{eqnarray}
u_i^2  &:=& 2 \left[ 1 - \cos(k_i \Delta x) \right] \; , \\
s_{ij} &:=& - \sin(k_i \Delta x) \sin(k_j \Delta x) \; .
\end{eqnarray}

Let us now define
\begin{equation}
\lambda := \frac{2}{\rho^2} \left[ 1 - \cos(\omega \Delta t)
\right] \; .
\end{equation}
Equation~(\ref{eq:FDA}) now becomes
\begin{equation}
\tilde{M} \hat{\bf h} = \lambda \hat{\bf h} \; ,
\end{equation}
which is just an eigenvalue equation.  Here we face one problem: the
characteristic polynomial is of 6th order, and is difficult to solve
exactly in the general case.  We will then consider a couple of
particular cases.

First, assume that the wave moves only on the $x$ direction, so $k_y =
k_z = 0$.  In this case everything simplifies considerably, and we find
that the eigenvalues of $\tilde{M}$ are just

\begin{itemize}

\item $\lambda = 0$, with multiplicity 3.

\item $\lambda = u_x^2$, with multiplicity 3.

\end{itemize}

This has precisely the same structure we found before for the exact
system of differential equations.  The only difference being that the
wave speed is now not quite 1.  The wave speed in fact depends on the
wave number $k_x$, and can be obtained from the dispersion relation
\begin{equation}
\frac{2}{\rho^2} \left[ 1 - \cos(\omega \Delta t)
\right] =  u_x^2 \; .
\end{equation}

Notice that for small $k_x \Delta x$ (large wavelengths compared to
the grid spacing) this relation reduces to
\begin{equation}
\omega^2 = k_x^2 \; ,
\end{equation}
which is what one expects.  For smaller wavelengths we obtain wave
speeds that are smaller than 1, showing the dispersive nature of the
finite difference approximation.

The results above are not particularly surprising.  One obtains
essentially the same thing for the simple wave equation.  The
interesting case is when we consider waves moving in a direction
different from the coordinate lines.  We will then consider the
particular case of waves moving in the diagonal direction, for which
$k_x = k_y = k_z \equiv k$.  The characteristic polynomial now does
not simplify nearly as much, but one can still find the eigenvalues
analytically.  They are

\begin{itemize}

\item $\lambda = u^2 - s^2$, with multiplicity 2.

\item $\lambda = u^2 + 2 s^2$, with multiplicity 2.

\item $\lambda = \frac{1}{2} \left[ 5 u^2 - 2 s^2 \pm \left( 9 u^4 - 12
s^4
+ 12 u^2 s^2 \right)^{1/2} \right]$.

\end{itemize}

\noindent where

\begin{equation}
u^2 = 2 \left[ 1 - \cos(k \Delta x) \right] , \qquad s^2 =
\sin^2(k \Delta x) .
\end{equation}

\begin{figure}
\epsfxsize=3.4in
\epsfysize=3.4in
\epsfbox{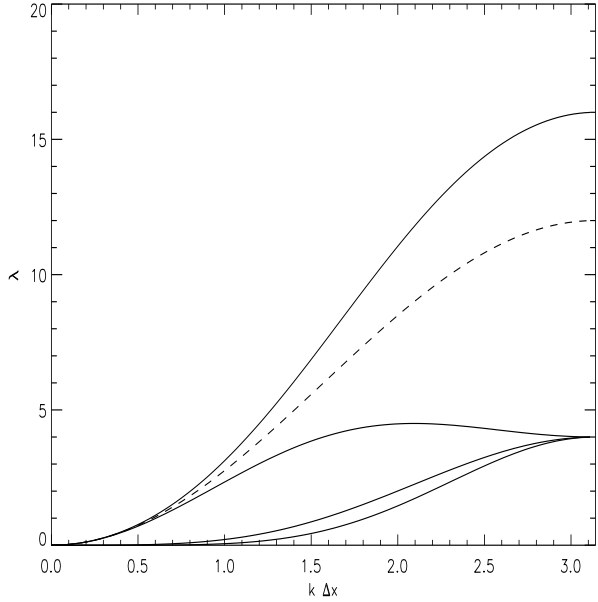}
\caption{Eigenvalues of the characteristic matrix $\tilde M$.  The
solid lines indicate the four distinct eigenvalues, while the dashed
line indicates the eigenvalue one would obtain for the finite
difference approximation to the simple 3D wave equation.}
\label{fig:roots}
\end{figure}

The values of the different roots are shown in Figure~\ref{fig:roots}.
The solid lines indicate the four distinct eigenvalues, while the
dashed line indicates the eigenvalue one would obtain (also along the
diagonal line) for the finite difference approximation to the simple
3D wave equation $\lambda = 3 u^2$.  The plot is only in the region $k
\Delta x \in [0,\pi]$ since larger wave numbers can not be represented
on the numerical grid ($k=\pi/\Delta x$ is the so-called `Nyquist'
frequency of our grid).

There are several things to notice from this result.  First, we now
have four distinct eigenvalues instead of two: the numerical grid has
broken the degeneracy of the exact problem.  Second, the three
eigenvalues that where zero in the exact case are now only zero for
$k=0$, and are clearly non-zero for any finite $k$.  This shows that
the zero speed modes have picked up a non-zero speed in the numerical
approximation.  This artificial speed is very small for large
wavelengths (small $k$), but becomes considerable for smaller
wavelengths.  Finally, we see that for small values of $k$ we recover
the exact result: one eigenvalue vanishes as $k^6/12$, two as $k^4/4$,
and the other three go to zero as $3 k^2$, which is the correct result
for waves traveling with a speed of 1 along the diagonal.


\bibliographystyle{prsty}
\bibliography{bibtex/references}

\end{document}